\documentclass[%
reprint,
superscriptaddress,
 amsmath,amssymb,
 aps,prx
]{revtex4-2}

\usepackage{graphicx}
\usepackage{dcolumn}
\usepackage{bm}
\usepackage[colorlinks=true,linkcolor=blue,urlcolor=blue,citecolor=blue]{hyperref}
\usepackage{ragged2e}

\usepackage{subfigure}
\usepackage{xcolor}
\usepackage{amsmath}
\usepackage{adjustbox}
\usepackage{physics}
\usepackage{float}
\usepackage{tikz}
\usepackage{graphicx}
\usepackage{dcolumn}%

\usepackage{rotating}
\usepackage{makecell}
\usepackage{booktabs}
\newcolumntype{V}[1]{>{\centering\arraybackslash}m{#1}}  
\newcolumntype{C}[1]{>{\centering\arraybackslash}p{#1}}  

\usepackage{caption}
\usepackage{adjustbox}
\usepackage{colortbl}
\usepackage{array}
\usepackage{tikz}
\usepackage{multirow}

 \graphicspath{ {./Images/} }
\begin{document}

\preprint{APS/123-QED}
\title{Bosonic Diffusive Channel: Quantum Metrology via Finite Non-Gaussian Resource}

\author{Arman}
\email{a19rs004@iiserkol.ac.in}
\affiliation{Department of Physical Sciences, Indian Institute of Science Education and Research Kolkata, Mohanpur-741246, West Bengal, India}

\author{Prasanta K. Panigrahi}%
\email{pprasanta@iiserkol.ac.in}
\affiliation{Department of Physical Sciences, Indian Institute of Science Education and Research Kolkata, Mohanpur-741246, West Bengal, India}

\date{\today}


\date{\today}

\begin{abstract}
    We investigate the estimation of dephasing-induced decoherence in continuous-variable quantum systems using non-Gaussian probe states. By purifying the open system, we identify optimal probes—specifically squeezed cat and symmetric squeezed compass states—via quantum Fisher information (QFI). These results are in agreement with numerical simulation.  In settings where the intra-cavity field is inaccessible and standard measurements are impractical, utilizing an ancilla approach where a qubit traverses or interacts with the cavity field, leading to measurement of the qubit, hence allowing estimation of the dephasing rate via Wigner function reconstruction or less costly marginal distribution.
\end{abstract}


\maketitle
\section{Introduction}
Metrology, a cornerstone of modern physics, plays a vital role not only in technological advancements but also in refining measurement strategies to achieve ever-greater precision. In the presence of statistical noise or, more generally, stochastic dynamics, the output of a parameter-dependent process exhibits inherent randomness, introducing uncertainty in the estimation of the parameter of interest. It is commonly assumed---though not strictly necessary---that the probe is sufficiently well-prepared such that its own stochasticity does not contribute significantly to the estimation error. This assumption typically holds in classical settings, where both the parameter and its fluctuations arise from the same physical process of interest. For example, a laser-cooled nanoparticle~\cite{levito} or micro-cluster~\cite{pahi2025} trapped in an optical potential exhibits Brownian motion~\cite{diffusion_einstein,einstein1987collected}, with a measurable mean position and variance. In such systems, the estimation of parameters like the diffusion constant or drift---appearing in the Fokker--Planck equation~\cite{walls_milburn_quantum_optics_2008,carmichael1999statistical,louisell1973quantum}---or a coherent displacement (e.g., a small kick of amplitude \(\mathbf{x}\)) involves analyzing observables such as the second moment \(\langle x^2 \rangle\). The uncertainty in estimating the parameter \(\mathbf{x}\) is inherently tied to the stochastic nature of the system, which may arise from random perturbations rather than a specific mechanism. The repeated measurements over \( \mathbb{N} \) trials typically yield Gaussian-distributed outcomes, with the mean and standard deviation scaling as \(\mu\) and \(\sigma/\sqrt{\mathbb{N}}\), respectively, according to the central limit theorem. This defines the \emph{shot-noise limit} (SNL), the classical bound on measurement precision. While classical correlations and optimized probes can marginally improve precision, surpassing the SNL and reaching \emph{sub-shot-noise}~\cite{mac_squeezing_science} sensitivity fundamentally requires the use of quantum features in the probe, the system, and the measurement protocol.

The interesting parameter precision in some cases depends on the observable, and in other cases on the choice of basis or space of variables as well, where the output is taken. The choice of basis comes into the picture more nicely in the context of quantum processes, where types of measurements also affect the error in the parameter. In any case, the error width from experimental data is extracted from the observable, depending on the variable where the stochastic nature arises, and this manifests itself in the measurement extraction of the parameter.

Quantum metrology has its relation to enhancing the precision of measurement, using quantum effects, for the estimation of parameters involved in quantum physical processes. The main quantum features that have been useful for quantum metrology are quantum superposition, squeezing~\cite{mac_squeezing_science}, and entanglement~\cite{mac_squeezing_science}. These quantum resources have properties describing coherence and correlation in the probes, leading to a sensitivity response for tiny changes in the parameters due to small magnitude coherent drive or external noise. In some cases, this sensitivity response can be seen from the phase-space structure of the quantum probe, such as small displacements~\cite{zurek} and phase shifts, specifically phase-space rotation~\cite{dalvit}. The sensitivity in quantum metrology is described as Fisher information in the Cramer-Rao bound~\cite{cavesgeometry}. This sensitivity behavior is extracted in the measurement of observables depending on the parameter to be estimated, such as fidelity, position, Wigner, or phase-space distribution. The maximum response depends on the choice of basis, leading to the specific selection of the observable with a non-zero expectation value~\cite{prlmaccone,cavesgeometry}.   

Either it is a coherent displacement ($s$)~\cite{armancompass, zurek, dalvit} due to external force~\cite{luis} or an incoherent displacement ($s$) because of noise~\cite{prlmaccone} in the phase-space, this effective shift $s$ is the parameter that holds information of the coherent force or external noise. This displacement measurement reveals information about the damping rate~\cite{prragerwal} that describes noise, other than the coherent force~\cite{scienceburd, ruppert2022high, preskill}. Similarly to classical estimation processes, as discussed above, the observable $\hat{O}_{\mathrm{t}}$ serving as a sensitive quantity, which changes sharply with a small parameter $\mathrm{t}$ or effective noise, leads to their precise estimation; however, this is highly affected by the quantum nature of a probe and measurement basis. 
Quantitatively, the sensitivity is given as $|\partial_{\mathrm{t}}\langle\hat{O}_\mathrm{t}\rangle|/\sqrt{\langle\Delta^{2} \hat{O}_{\mathrm{t}}\rangle}$, maximally bounded by QFI ($F_{Q}$). Apart from extensive work on the estimation of parameters in the absence and presence of noisy channels~\cite{prlmaccone, preskill}, little work has been performed on the metrology of decoherence rates, specifically amplitude~\cite{prragerwal} and dephasing channels~\cite{PRXQuantumWilde}. There has been discussion with Gaussian resource~\cite{PRXQuantumWilde} states providing certain precision in diffusion or bosonic dephasing, which we hope to enhance with non-Gaussian~\cite{jointdiffusion} resources as an indicator of non-classicality~\cite{PRL-jeong, shor}. It is worth noting that such resources have been utilized in amplitude rate estimation~\cite{prlmaccone, armancompass}, proving to be useful in improving precision.





In this paper, we describe the dephasing channel and illustrate the phase-space behavior (Sec.\hyperref[sec:II]{II}) for three states: coherent, squeezed, and cat state. Furthermore, we utilize the purification process for the same channel discussed in the previous section to gain Fisher information in sec.\hyperref[sec:III]{III} with a description of non-Gaussian states. At last, we obtain numerically exact quantum Fisher information (QFI) via symmetric Logarithmic derivative (SLD) in  Sec.\hyperref[sec:IV]{IV}. Finally, Sec.\hyperref[sec:V]{V} contains a conclusion with a summary.

\section{Phase Space Diffusive Channel}\label{sec:II}

In open quantum systems, intra-cavity light undergoes two primary forms of decoherence: photon loss and dephasing. Photon loss reduces both quantum coherence (as quantified by the \( l_1 \)-norm) and the average photon number \( \langle \hat{n} \rangle \), typically arising from beam splitters, partially transmitting cavity mirrors, or Jaynes–Cummings (JC)–type system-bath interactions. This process has been widely studied in quantum metrology, particularly for estimating cavity damping rates and beam-splitter transmissivities.

 It induces small displacements in phase space in the infinitesimal-time limit \cite{Gri-gkp}. This process is described by the Kraus operator representation of the form  
\[
\rho(t) = \int_{-\infty}^{\infty} d^2\alpha\, p(\alpha) D[\alpha] \rho D^\dagger[\alpha],
\]  
which models partial coupling to the environment and solves a heat equation that differs from the standard thermal bath model \cite{Gri-gkp}. This description accounts for both photon loss and gain with symmetric decoherence rates. More generally, bosonic channels can be modeled by displacement Kraus operators as  
\[
\rho(t) = \int_{-\infty}^{\infty} d^2\alpha\, d^2\beta\, p(\alpha, \beta) D[\alpha, \beta] \rho D^\dagger[\alpha, \beta],
\]  
capturing both amplitude damping and dephasing processes \cite{Gri-gkp}.
In contrast, dephasing degrades coherence while preserving the average photon number. Specifically, dephasing is modeled by an ensemble of phase-space rotations using operators \( e^{i\hat{n}\phi} \), giving rise to  
\[
\rho(t) = \int_{-\infty}^{\infty} d\phi\, p(\phi)\, e^{i\hat{n}\phi} \rho_0\, e^{-i\hat{n}\phi},
\]  
with  
\[
p(\phi) = \frac{1}{\sqrt{2\pi \kappa_{\mathrm{dph}} t}} e^{-\phi^2 / (2\kappa_{\mathrm{dph}} t)},
\]  
representing the probability distribution over phase shifts \cite{Gri-rot, prlmaccone}. This describes a mixture of all states resulting from phase rotations, with larger dephasing constants enabling broader access to the phase space and thus faster randomization. The evolution under dephasing is typically slower than amplitude damping, \( \kappa_{\mathrm{dph}} t \ll \kappa_d t \), due to weak coupling between the system and the bath.

The Wigner function and associated phase-space distributions reveal the diffusive nature of dephasing channels, particularly when the initial states are (i) coherent states \( \ket{\alpha} \), (ii) squeezed states \( \ket{\zeta} \), and (iii) Schrödinger cat states \( \ket{\alpha} + \ket{-\alpha} \). Under dephasing, these initially structured distributions evolve toward circularly symmetric profiles, indicating equal phase probability.

This behavior is captured by the angular diffusion equation in phase space:

\[
\dot{W}(|\beta|e^{i\theta}) = \kappa_{\mathrm{dph}} \partial_\theta^2 W(|\beta|e^{i\theta}),\]
\[\text{and} \quad \dot{\mathbb{P}}_{W}(\theta) = \kappa_{\mathrm{dph}} \partial_\theta^2 \mathbb{P}_{W}(\theta),
\]
where the Wigner function is defined as \( W[\beta] = \frac{2}{\pi} \Tr[\rho\, D[2\beta]\hat{\Pi}] \), and the phase probability distribution is given by  
\[
{\mathbb{P}}_W(\theta) = \frac{1}{2} \int_0^\infty d|\beta|^2\, W(|\beta|e^{i\theta}).
\]

\begin{figure}[tb]
\centering
\begin{tikzpicture}
\node[fill=cyan!10!white, draw=green!5, line width=0.6pt, rounded corners=4pt, inner sep=6pt] (box) {
\begin{adjustbox}{width=0.95\columnwidth}
\begin{tabular}{V{0.5cm} C{0.25\linewidth} C{0.25\linewidth} C{0.25\linewidth}}
 & \textbf{$\kappa_{\text{dph}} t=0.05$} & \textbf{$\kappa_{\text{dph}} t=0.1$} & \textbf{$\kappa_{\text{dph}} t=1$} \\
\rotatebox{90}{\hspace{30pt}$\mathbf{\ket{\alpha}}$\hspace{5pt}p} &
\includegraphics[width=\linewidth]{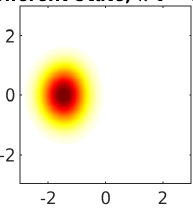} & 
\includegraphics[width=\linewidth]{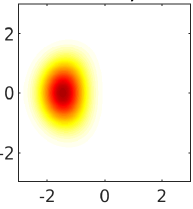} & 
\includegraphics[width=\linewidth]{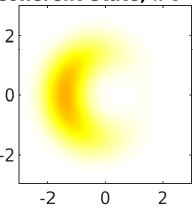} \\
\rotatebox{90}{\hspace{30pt}$\mathbf{\ket{\zeta}}$} &
\includegraphics[width=\linewidth]{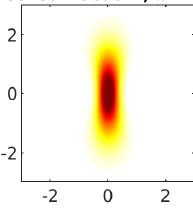} & 
\includegraphics[width=\linewidth]{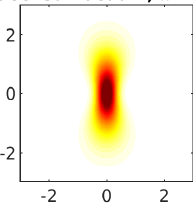} & 
\includegraphics[width=\linewidth]{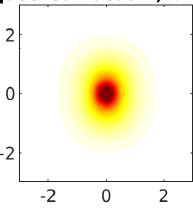} \\
\rotatebox{90}{\hspace{16pt}$\mathbf{\ket{\alpha}+\ket{-\alpha}}$} &
\includegraphics[width=\linewidth]{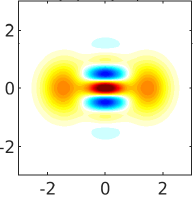} & 
\includegraphics[width=\linewidth]{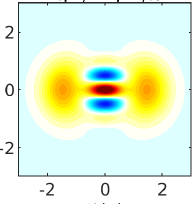} & 
\includegraphics[width=\linewidth]{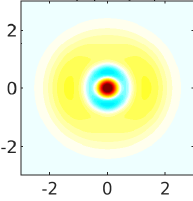} \\
 & \multicolumn{3}{c}{
     \includegraphics[width=0.8\linewidth]{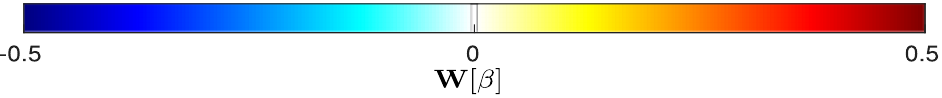}
   } \\
\end{tabular}
\end{adjustbox}
};
\end{tikzpicture}

  \caption{\justifying (Color online) The qualitative illustration of the dephasing channel via phase space diffusion of Gaussian and non-Gaussian states, when initialized in the cavity, subjected to the same dephasing induced-decoherence. The phase-space distribution ($W[\beta$]) for coherent state $\left(\ket{\alpha}\right)$ (Upper Row), squeezed state $\left(\ket{\zeta}\right)$ (Middle Row), and cat state $\left(\ket{\alpha}+\ket{-\alpha}\right)\,$(Lower Row) are shown for increasing decoherence time ($t$), leading to the symmetrical distribution around origin along phase ($\theta$).}
    \label{fig:illus_wigner}
\end{figure}

These dynamics are equivalent to angular diffusion akin to the Brownian motion \cite{diffusion_einstein,einstein1987collected}, with the phase-space distribution spreading along the angular coordinate \( \theta \) under dephasing as seen in the Fig.\ref{fig:illus_wigner}. 


\section{Purification via Opto-Mechanical Interaction}\label{sec:III}

The same channel for the single-mode field with $\kappa_{\text{dph}}$ dephasing rate, characterizing  phase-space diffusion-induced decoherence, is represented as
\begin{equation}
    \frac{\partial\rho}{\partial t}=\frac{\kappa_{\text{dph}}}{2}\left(2\hat{n}\rho \hat{n} - (\hat{n})^{2}\rho -\rho (\hat{n})^{2}\right).
    \label{Eq:diffusion}
\end{equation}
Similar coherent different processes, where the initial phase-space distribution disperses symmetrically and circularly in the phase-space along $\theta$ directions, are due to Kerr-type interactions ($\hat{a}^{\dagger\,2}\hat{a}^{2}$)~\cite{TarafracRevival} realizable in optical, ion-trapping and superconducting platforms as well as optomechanical $\hat{a}^{\dagger}\hat{a}(\hat{b}^{\dagger}+\hat{b})$ type interactions~\cite{Quantum_Hypercube_States}. These coherent dynamics not only have similar behavior to quantum phase diffusion, but also produce negative regions in the phase space, evolving from a positive initial Wigner distribution. Interestingly, any initial state in this dephasing-induced decoherence channel retains negative regions in phase space, thus still possesses non-classicality. Using thermo-entangled state representation, Eq.~\ref{Eq:diffusion} can be solved for the density matrix in terms of Kraus operator representations as

\[
\rho(t) = \sum_{k=0}^{\infty} \frac{(\kappa_{\text{dph}} t)^k}{k!} \hat{n}^k e^{-\frac{\kappa_{\text{dph}} t}{2} \hat{n}^2} \rho_0 e^{-\frac{\kappa_{\text{dph}} t}{2} \hat{n}^2} \hat{n}^k.
\]

\begin{center}
\begin{figure}[tb]
    \centering\includegraphics[width=\linewidth]{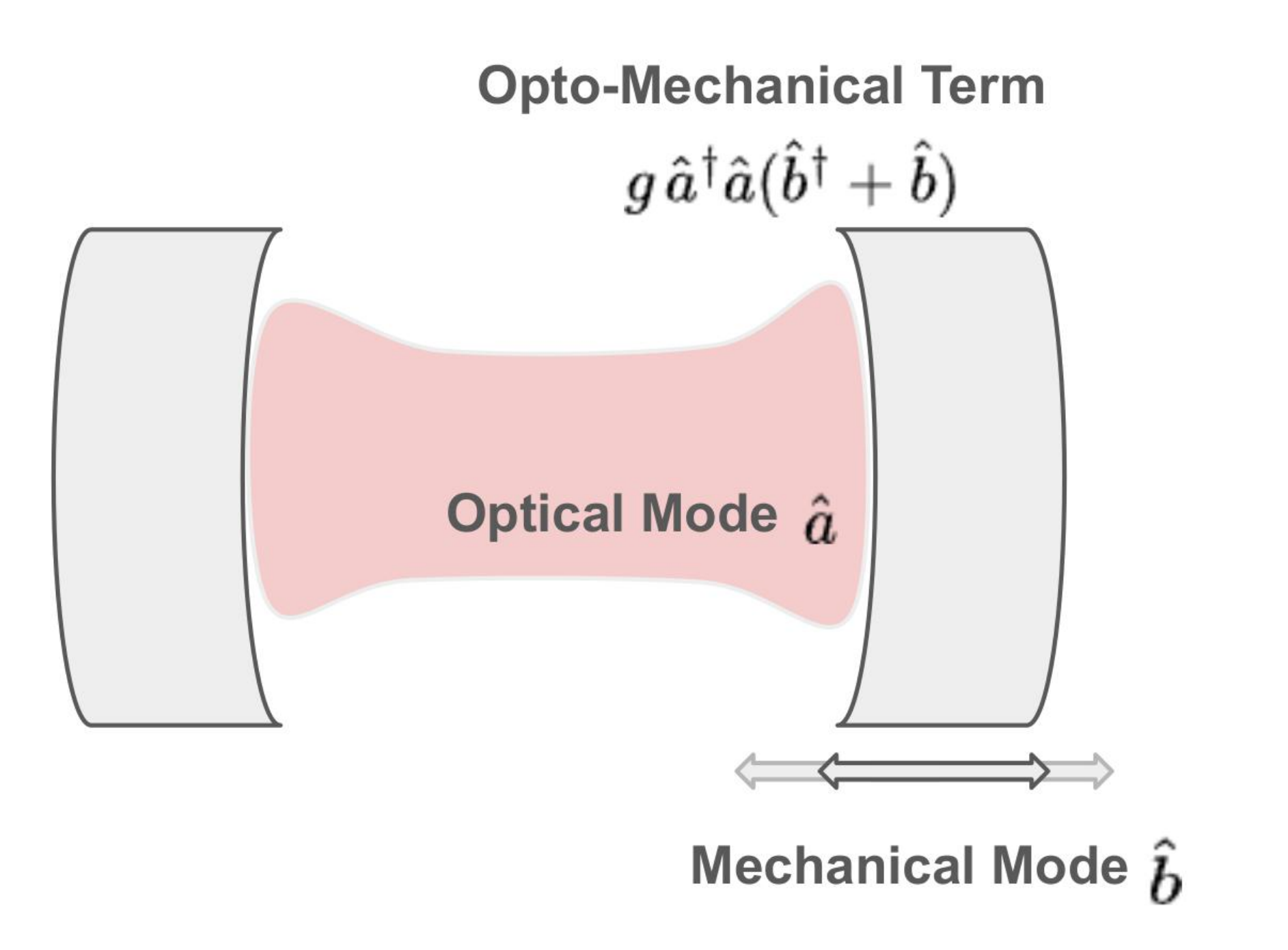}
    \caption{\justifying (Color online) Illustration of the dephasing-induced decoherence channel under purification via single-mode bath interactions, specifically opto-mechanical coupling. An alternative form of bath interaction can also be considered for purification or generalized modeling by incorporating the spectral response of the bath to the system via virtual bath excitation, thereby capturing diffusion-induced decoherence.}
    \label{fig:opto_mechanical}
\end{figure}
\end{center}

This density matrix can be purified with the help of the optomechanical interaction present in the cavity between the mechanical oscillator (mirror) and the cavity field. We perform the evolution of the initial cavity field ($\ket{\psi}_{c}$) and mechanical oscillator ($\ket{0}_{m}$) state with unitary operator $U=e^{i \mathcal{H} t}$ as follows

\begin{multline*}
    \ket{\psi(t)} =e^{-it(\omega_{c}a^{+}a+\omega_{m}b^{+}b)}e^{\frac{-|\lambda|^{2}(a^{+}a)^{2}}{2}}\\\sum_{k=0}^{\infty}\frac{(-i\lambda^{*})^{k}}{k!}(a^{+}ab^{+})^{k}e^{-i\lambda a^{+}ab}\ket{\psi_{o}},
\end{multline*}
where parameter $\lambda=gtj_{o}(\omega_{m}t/2)e^{-i\omega_{m}t/2}$ and system Hamiltonian, $$\mathcal{H}=\omega_{c}a^{\dagger}a+\omega_{m}b^{\dagger}b+ g a^{\dagger}a(b^{\dagger}+b).$$   

This type of dephasing can appear from bath interaction, either through an optomechanical type interaction where the bosonic mode of the system interacts with the mechanical position mode $(b^{\dagger}+ \hat{b})$, or a bath virtual excitation of type similar to $\hat{b}^{+}\hat{b}$ \cite{Blais, prlsavasta}.
Assuming many-modes interaction of bath ($\hat{b}$) $\,\mathcal{H}=\omega_{c}a^{+}a+\sum_{m=0}^{\infty}(\omega_{m}b_{m}^{+}b_{m}+ g_{m} a^{+}a(b_{m}^{+}+b_{m}))$ coupled with the system should lead to generalized spectrum inculsive to dephasing rate $\kappa_{\text{dph}}$; however it has been considered on one mode for the purification process, such as for the beam splitter\cite{prragerwal, prrfisher, scienceburd} and phase~\cite{GaussianQND}. The cavity field evolution is obtained for opto-mechanical interaction by tracing out the mechanical mode in the field and the mechanical oscillator interaction with $\ket{\psi_{o}}=\ket{\psi_{c}}\ket{0}_{m}$ as follows 


 $$\rho_{c}(|\lambda|)= \mathcal{\textnormal{Tr}}_{m}\left[\ket{\psi(t)}\bra{\psi(t)}\right] $$
 $$  =\sum_{k=0}^{\infty}\frac{|\lambda|^{k}}{\sqrt{k!}}(\hat{n})^{k}e^{\frac{-|\lambda|^{2}(\hat{n})^{2}}{2}}\rho_{c}(0)e^{\frac{-|\lambda|^{2}(\hat{n})^{2}}{2}}(\hat{n})^{k}\frac{|\lambda|^{k}}{\sqrt{k!}}.$$

We use a unitary transformation ($U$) due to optomechanics to obtain the generator ($G$) of dephasing ($\kappa_{d}$) in the system of the cavity field. The generator $G$ is given by $$\hat{G}=-i\frac{\partial \hat{U}}{\partial g}\hat{U}^{\dagger}=-tj_{0}(\omega_{m}t/2)a^{+}a(be^{-i\omega_{m}t/2}+b^{+}e^{i\omega_{m}t/2}).$$
Write about Fisher information and how it can be related to observable in the experiments. 
This generator ($\hat{G}$) provides quantum Fisher information ($F_{Q}^{\mathit{p}}$) in terms of the variance
$$F_{Q}^{\mathit{p}}=\langle\Delta^{2}\hat{G}\rangle =\left(tj_{0}(\omega_{m}t/2)\right)^{2}\langle\hat{n}^{2}\rangle$$
$$=\left(tj_{0}(\omega_{m}t/2)\right)^{2}\left( \langle\Delta^{2}\hat{n}\rangle+\langle \hat{n}\rangle^{2}\right).$$
Similarly, $F_{Q}^{\mathit{p}}$ is expressed for $|\lambda|$ under purification as $\langle\hat{n}^{2}\rangle$.
The Quantum Cramer-Rao bound for estimating the coupling strength $g$, leads to
$$(\delta g)^{2}\geq
\left[\left(t\,j_{0}(\omega_{m}t/2)\right)^{2}\langle \hat{n}^2\rangle\right]^{-1},$$
and again for parameter $|\lambda|$ analogous to the dephasing constant ($\kappa_{d}$)
$$(\delta|\lambda|)^{2}\geq
\left[\langle \hat{n}^2\rangle\right]^{-1}.$$

\begin{center}
\begin{figure}[htbp!]
    \centering\includegraphics[width=1\linewidth]{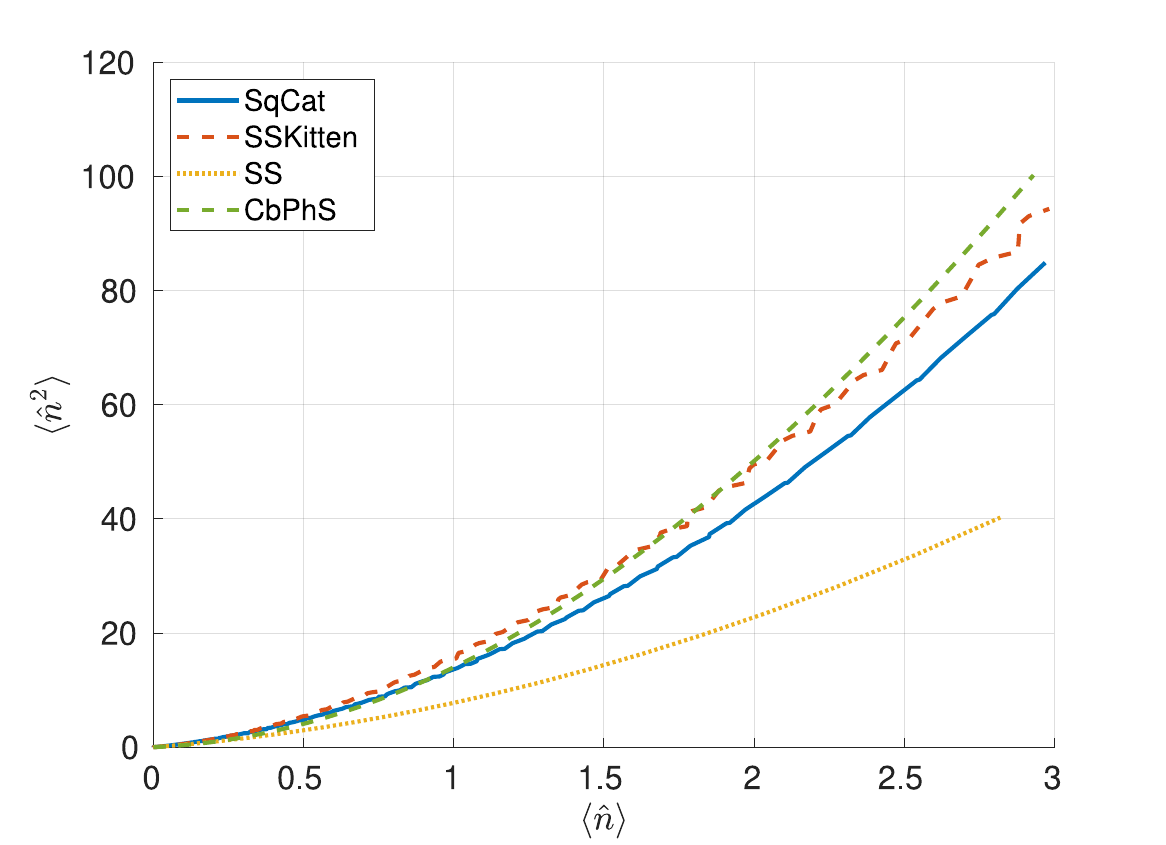}
    \caption{\justifying (Color online) Obtained Fisher information $\langle\hat{n}^{2}\rangle$ vs. average energy $\langle\hat{n}\rangle$ of non-Gaussian probes for quantum diffusive channel when purified via single-mode opto-mechanical type interactions, considering the mechanical bath is initialized at vacuum. The symmetric squeezed compass state (SSKitten) performs well in the regime of small energy $0<\langle\hat{n}\rangle<1.5$ as compared to the other considered non-Gaussian probes: cubic phase state (CbPhS), squeezed cat state (SqCat), and symmetric squeezed state (SS).}
    \label{fig:purified_fisher}
\end{figure}
\end{center}

From this result, we find that deviation associated in estimating $g$ (optomechanical interaction strength $g$ is mimicking dephasing constant $\kappa_{d}$ in the cavity system) is dependent on both aspects, the average photon energy $\langle \hat{n}\rangle$ and Agarwal's $Q$ parameter $\left(\Delta^{2}\hat{n}/\langle\hat{n}\rangle\right)$ statistics of the probe state. This number variance ($\Delta^{2}\hat{n}$) dependence for the Fisher information is also seen in the phase shift estimation when light passes through phase-changing medium \cite{phaseshit0,phaseshit1,prrfisher,scienceburd,phaseshit4} and diffusive channels \cite{GaussianQND,jointdiffusion}. By the relation of $F_{Q}^{\mathit{p}}$, it is apparent that this is maximized when the number variance is equal to the average number square, essentially signifying super-Poissonian statistics or bunching nature of a pure state as a probe. These behaviours are seen in the simple and minimal non-classical pure states such as coherent and squeezed states. Other possible probe states for the $Q$ parameter greater than $1$, are the even superposition of squeeze and displacement ones, which include vacuum with a large weight factor in the Fock space expansion. We use multiple squeezed and coherent states superpositions (see Table.~\ref{tab:nomenclature}) to investigate their statistics, whether there exists a super-Poissonian $Q>1$ regime. 

For sake of completeness, some important operators and states considered in the Table.~\ref{tab:nomenclature} are displacement $D[\alpha]=e^{\alpha \hat{a}^{\dagger} -\alpha^* \hat{a}}$, squeeze $S[2\zeta]=e^{\zeta^* \hat{a}^{2} -\zeta \hat{a}^{2}}$ and cubic phase $\Gamma[\gamma]=e^{-i\gamma\hat{x}^{3}}$ operators following coherent state $\ket{\alpha}=D[\alpha]\ket{0}$ and vacuum $\ket{0}$, the ground state of harmonic oscillator. The mod-SSW state is the modified phase state discussed in the literature~\cite{GaussianQND}.


\section{Numerical Quantum Fisher Information}\label{sec:IV}
This metrological potential can be further explored for the considered non-Gaussian states~\cite{NoDiffusion} using an exact analysis of the quantum Fisher information (QFI). In diffusive channels, the initial state's purity degrades over time, with the density matrix approaching a diagonal form in the Fock basis. As a result, deriving an analytical expression for the QFI becomes intractable, necessitating numerical methods for evaluation.

To address this, we compute the QFI numerically using the locus derived from the purification of the quantum state. The exact QFI for parameter-dependent processes is defined in terms of the symmetric logarithmic derivative (SLD)~\cite{cavesgeometry}:
\[
2\partial_{|\lambda|}\rho = \rho \hat{L} + \hat{L} \rho,
\]
where the SLD operator is given by
\[
\hat{L} = \sum_{k,l=0}^{\infty} \frac{2 \bra{k} \partial_{|\lambda|} \rho(|\lambda|) \ket{l}}{p_k + p_l} \ket{k} \bra{l},
\]
and the QFI is computed as \( F_Q = \Tr[\rho \hat{L}^2] \). The parameter-induced density matrix ($\rho$), in terms of probability ($p_{k}$), is given as $\rho_{|\lambda|}=\sum_{k=0}^{\infty}p_{k}\ket{k}\bra{k}.$

\begin{center}
\begin{figure}[htbp!] 
 \centering\includegraphics[height=0.7\linewidth,width=1\linewidth]{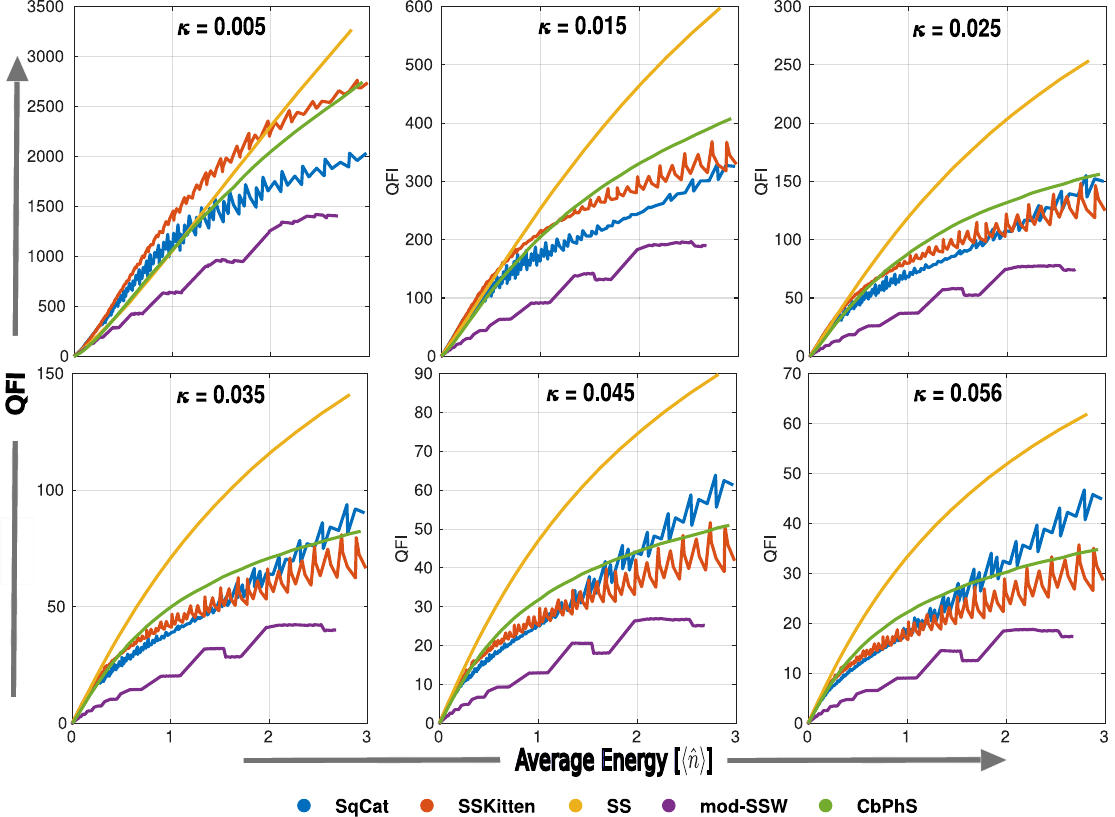}
 \caption{\justifying
Quantum Fisher Information (QFI) obtained numerically via the symmetric logarithmic derivative (SLD) operator \(\hat{L}\), plotted against the average photon number \(\langle \hat{n} \rangle\) for a dephasing channel. The non-Gaussian probe states—SSKitten and SqCat—demonstrate superior performance for local estimation in the low average energy regime, while the SS yields higher QFI at large dephasing rates, indicating stronger global sensitivity. These results underscore the advantage of non-Gaussian pure probes, which offer metrological gains comparable to or exceeding those of Gaussian states, but at finite energy cost.}

    \label{fig:numerical_fisher}
\end{figure}
\end{center}

Furthermore, the QFI is directly related to the Bures metric \( d_{\mathrm{Bures}}^2(\rho_1, \rho_2) \)~\cite{cavesgeometry, SqOptiDiffusion, HUBNERBures}, via:
\[
d_{\mathrm{Bures}}^2(\rho_{|\lambda|}, \rho_{|\lambda| + \delta|\lambda|}) = (\delta|\lambda|)^2 F_Q,
\]
\[
= \lim_{\delta|\lambda| \to 0} \, 8 \left(1 - \mathcal{F}(\rho_{|\lambda|}, \rho_{|\lambda| + \delta|\lambda|})\right),
\]
where \( \mathcal{F}(\rho_1, \rho_2) \) denotes the fidelity between the two density matrices \( \rho_1 \) and \( \rho_2 \).

We consider a combination of the Gaussian and non-Gaussian probes (see table~\ref{tab:nomenclature}), which suggests inherently localized distributions in the phase-space ($W\left[|\beta|e^{i\theta}\right]$) along angular dimensions ($\theta$), may lead to large phase diffusion as being sensitive to concentrated distributions in phase space.
\begin{table}[H]
\caption{\label{tab:nomenclature} Non-Gaussian states and their Nomenclature for the diffusive parameter estimations. Here, $i=\sqrt{-1}.$ }
\centering
\begin{ruledtabular}
\begin{tabular}{lcl}
Notation & State $\ket{\psi}$ & Name \\
\hline
Kitten & $\displaystyle \sum_{k=0}^{3}\ket{i^{k}\alpha}$ & Kitten/Compass State (KS) \\[1pt]
Cat & $\displaystyle \ket{\alpha} + \ket{-\alpha}$ & Cat State \\[1pt]
SqCat & $\displaystyle \sum_{k=0}^{1} S[r]\ket{i^{2k}\alpha}$ & Squeezed Cat State \\[3pt]
SS & $\displaystyle \sum_{k=0}^{1} S[i^{2k}r]\ket{0}$ & Symmetrically Squeezed State \\[3pt]
SqCS & $\displaystyle S[r]\ket{\alpha}$ & Squeezed Coherent State  \\[3pt]
SSKitten & $\displaystyle \sum_{l=0}^{1}S[i^{2l}r]\sum_{k=0}^{3}\ket{i^{k}\alpha}$ & Symmetrically Squeezed KS  \\[3pt]
CbPhS & $\displaystyle \Gamma[\gamma]\ket{0}$ & Cubic Phase State  \\[3pt]
mod-SSW & $\displaystyle \sum_{n=0}^{s}(n+q)^{-p}\ket{n}$ & modified-SSW State\\

\end{tabular}
\end{ruledtabular}
\end{table}

\section{conclusion}\label{sec:V}
We investigate a single-mode quantum \textit{dephasing channel}, in which an initially localized distribution in the angular (phase) dimension of phase space undergoes random smearing. This leads to a \textit{circularly symmetric distribution} in the steady state, resembling the Wigner function of a Fock state. Consequently, this channel is often referred to as a \textit{diffusive channel}. Notably, even though the quadrature distribution of a dephased quantum system may resemble that of a Fock state, the system remains \textit{nonclassical} due to the residual \textit{negativity in the Wigner function}.

While most prior studies have focused on the use of \textit{Gaussian states} for probing dephasing-induced diffusion~\cite{jointdiffusion, GaussianQND, PRXQuantumWilde}, here we explore a range of \textit{non-Gaussian and nonclassical quantum states} as probes. Through the purification approach, we demonstrate that the \textit{quantum Fisher information (QFI)} for the dephasing process takes the same analytical form as that obtained for coherent phase shifts in closed quantum dynamics. Specifically,
\begin{itemize}
    \item for dephasing channels, the QFI is proportional to \(\langle \hat{n}^2 \rangle\), 
    \item whereas for coherent unitary processes, it is proportional to the number variance \(\langle \Delta^2 \hat{n} \rangle\).
\end{itemize}

This insight guides the choice of \textit{non-Gaussian probes with high super-Poissonian statistics}, which are especially effective for metrological tasks involving decoherence. In our study, we consider a diverse set of such states. Among them, the \textit{squeezed kitten state (SSKitten)}—constructed by combining squeezing with multiple coherent-state superpositions—exhibits superior performance in both the analytically purified (see Fig.~\ref{fig:purified_fisher}) and numerically simulated scenarios (see Fig.~\ref{fig:numerical_fisher}), compared to other states such as the \textit{squeezed even cat} and the \textit{even cat} state, both of which are also known to possess super-Poissonian statistics.

For future applications, the \textit{observable quantities} that are suitable for estimating the parameter of interest—namely, the \textit{diffusion rate}—can be identified by evaluating the \textit{classical Fisher information (CFI)}. Unlike the quantum Fisher information \(F_Q\), which is measurement-independent and provides the ultimate precision bound, the CFI depends on the choice of measurement \textit{basis} (POVM: $\{\Pi_i|\sum_{i}\Pi_{i}=\mathbb{I}\}$). When the chosen basis closely approximates the optimal one, the CFI approaches the QFI, allowing one to identify observables (operators) that contain this optimal basis.

For instance, if the position basis \(\{\Pi =\ket{x}\bra{x}\}\) or probe itself \(\{\Pi =\ket{\psi}\bra{\psi}\}\)~\cite{prlmaccone} optimizes the CFI, then the observables best suited for estimating the parameter will be functions of the position operator, i.e., \(f(\hat{x})\).

The \textit{biased classical Fisher information}~\cite{cavesgeometry} can be determined using the \textit{error propagation formula}:
\[
\delta \kappa = \frac{\Delta f(\hat{x})}{\left| \partial_{\kappa} \langle f(\hat{x}) \rangle \right|},
\]
where the function \(f(\hat{x})\) is chosen to \textit{minimize} the uncertainty \(\delta\kappa\), ideally such that \(\delta\kappa \rightarrow 0\), thereby optimizing parameter sensitivity.

With recent advances in quantum information processing technologies, the preparation of highly \textit{non-Gaussian states} has become increasingly feasible—particularly on \textit{superconducting circuit} and \textit{trapped ion} platforms. These systems offer fine-tuned control over interaction strengths and coupling mechanisms, which makes them more amenable to engineering such states compared to traditional \textit{optical platforms}.

In this context, the generation of \textit{compass states}, especially in the presence of squeezing or in superpositions involving squeezed components, demands further investigation. A comparative study across different experimental platforms is essential to evaluate their practical feasibility and performance for metrological and foundational applications.


\begin{acknowledgements}
Arman is thankful to the University Grants Commission and Council of Scientific and Industrial Research, New Delhi, Government of India for Junior Research Fellowship at IISER Kolkata.
\end{acknowledgements}

\appendix

 \bibliography{biblio.bib}

\end{document}